\documentclass[12pt]{spieman}  
\usepackage{amsmath,amsfonts,amssymb}
\usepackage{graphicx}
\usepackage{setspace}
\usepackage{tocloft}
\usepackage{caption}
\usepackage{subcaption}
\usepackage[labelfont=bf]{caption}
\usepackage{lineno}
\usepackage[normalem]{ulem}

\title{Soft x-ray detection for small satellites with a commercial CMOS sensor at room temperature}

\author[a,*]{Steve Tammes}
\author[a,*]{Tyler Roth}
\author[a]{Philip Kaaret}
\author[a]{Casey DeRoo}
\author[a]{Abdallah Elmaleh}
\author[b]{Jessica L. McChesney}
\author[b]{Fanny Rodolakis}
\affil[a]{Department of Physics and Astronomy, University of Iowa, Iowa City, IA 52242, USA}
\affil[b]{Argonne National Laboratory, 9700 South Cass Avenue, Argonne, Illinois 60439, USA}

\cftpagenumbersoff{figure}
\cftpagenumbersoff{table}

\begin{document} 
\maketitle

\begin{abstract}
Recently CMOS (complementary metal-oxide-semiconductor) sensors have progressed to a point where they may offer improved performance in imaging x-ray detection compared to the CCDs often used in x-ray satellites. We demonstrate x-ray detection in the soft x-ray band (250-1700 eV) by a commercially available back-illuminated Sony IMX290LLR CMOS sensor using the Advanced Photon Source at the Argonne National Laboratory. While operating the device at room temperature, we measure energy resolutions (FWHM) of 48 eV at 250 eV and at 83 eV 1700 eV which are comparable to the performance of the \emph{Chandra} ACIS and the \emph{Suzaku} XIS. Furthermore, we demonstrate that the IMX290LLR can withstand radiation up to 17.1 krad, making it suitable for use on spacecraft in low earth orbit. 
\end{abstract}

\keywords{CMOS Image sensor, Back-side illuminated, {x-rays} spectroscopy, x-ray Imaging, Silicon, Radiation Effects, Sensor performance}

{\noindent \footnotesize\textbf{*}Steve Tammes,  \linkable{steven-tammes@uiowa.edu} }\\
{\noindent \footnotesize\textbf{*}Tyler Roth,  \linkable{tyler-roth@uiowa.edu} }

\begin{spacing}{2}   

\section{Introduction}
\label{sect:intro}  

The availability of low-cost small satellites for astrophysics drives a need for low-cost detectors that place minimal demands on the limited spacecraft resources available\cite{kaaret2014halosat}. The charge-coupled device (CCD) detectors often used for {x-ray} detection on space-based x-ray observatories have high cost and require cooling to temperatures near -100~K.\cite{turner2001european,garmire2003advanced,lamarr2004ground}  Complementary Metal-Oxide-Semiconductor (CMOS) sensors are designed for operation at room temperature and many commercial devices are available at low cost.  The active pixel design of CMOS sensors enables lower power consumption, faster readout rates, and comparable noise levels to CCDs\cite{bigas2006review}.

Here, we investigate the candidacy of a low-cost commercial CMOS sensor as an x-ray spectrometer in low earth orbit (LEO) for future small satellite missions. In Sec.~2, we describe the selected CMOS sensor, initial results on {x-ray} detection, and our {x-ray} event processing algorithm. We characterize the soft {x-ray} performance of the sensor using measurements obtained at an {x-ray} synchrotron beamline at the Argonne National Laboratory in Sec.~3. We then discuss the electron noise (Sec.~4), radiation testing (Sec.~5), and our conclusions (Sec.~6).

\section{{x-ray} Detection And Event Processing}
\label{sec:x-ray detection}
\subsection{{x-ray} Detection}

Back-illuminated (BI) sensors can provide superior performance for soft {x-ray} detection because the photons do not pass through the passivation, metallization, and inter-dielectric layers that form the pixel electronics. We evaluated several candidate BI sensors and selected the Sony IMX290LLR CMOS sensor for additional study due to its low electronic noise which is important in achieving good spectral resolution for soft {x-ray}s. The IMX290LLR contains a 1936$\times$1096 array of 2.9~$\mu$m square back-side-illuminated pixels and can achieve frame rates of up to 135 fps. To minimize development costs, we used an IDS Imaging Development Systems (https://en.ids-imaging.com/) UI-3860-LE-M-GL camera, that includes an IMX290 sensor and circuitry to provide a USB interface, and the associated Software Development Kit (SDK). For x-ray detection, we removed the coverglass designed to protect the sensor's surface.

\begin{figure}[tb]
\centering
  \centering
  \includegraphics[width=1.0\linewidth]{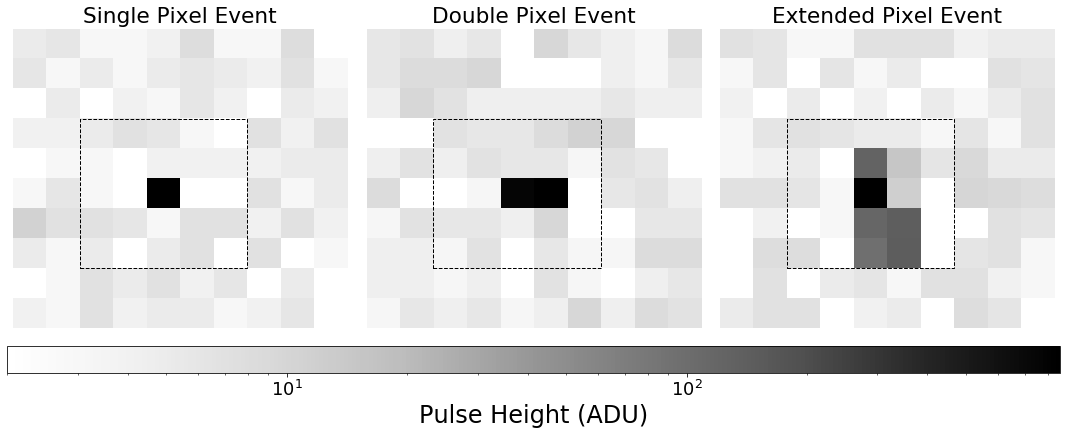}
\caption{Cutout images of the three types of events we consider for event processing. Shown from left to right are single pixel, double pixel, and extended pixel events. These events correspond to energies of 3.3 keV, 3.1 keV, and 5.4 keV respectively. }
\label{fig:xray_events}
\end{figure}

Our initial {x-ray} tests were performed using an $^{55}$Fe radioactive source and an {x-ray} tube fluorescing a teflon target and the walls of the vacuum chamber where the sensor was mounted. This produces characteristic emission lines from several elements including F, Cr, Mn, Fe, and bremsstraulung continuum radiation. The chamber was evacuated to $\approx 10^{-2}$ Torr and the sensor was mounted on a chiller plate maintained at a constant temperature of 21$^{\circ}$C. We acquired 4,000 images each with 300~ms exposure under {x-ray} illumination and 100 frames with no x-rays present for dark and bias measurements. Cutouts of x-ray events taken from single frames are shown in Fig \ref{fig:xray_events}. {x-ray}s produce events in which charge is deposited in one, two, or multiple pixels.

To select an appropriate gain for the sensor, we collected a series of spectra using the $^{55}$Fe radioactive source at different gain settings. {We measured the centroid of the pulse height distribution for the Mn~K-$\alpha$ emission line (corresponding to 1620 $e^{-}$ for an average hole-pair creation energy of $w = 3.64 \rm \, eV$ in silicon at room temperature \cite{scholze1996measurement}) using the event processing methodology outlined in Sec \ref{sec:event_processing}, and adopted settings producing a gain and associated fitting error of 0.991 $\pm$ 0.006 ADU/e$^-$}. These gain settings were used for all of the data described below.

\subsection{Event Processing}
\label{sec:event_processing}

Individual x-rays may deposit charge in multiple pixels due to the finite extent of the track traversed by primary photoelectron, the ejection of Auger electrons, and charge diffusion. To cope with the variable spread of charge across pixels, the event processing algorithms used for x-ray CCDs, such as Chandra's Advanced CCD imaging spectrometer (ACIS), search for pixels with charge greater than the `event threshold' and then record values in the surrounding 3$\times$3 pixel grid. Pixels above the `split pixel threshold' and sharing a side with the central pixel or in a 2$\times$2 square including the central pixel are included in the sum used to reconstruct the energy of the incident photon \cite{garmire2003advanced}. As shown in Fig.~\ref{fig:xray_events}, the {x-ray} events from the IMX290 sometimes extend over a larger range of pixels. This is likely due to the relatively small size of the pixels. We have developed an event processing algorithm utilizing a 5$\times$5 grid of pixels.


Before processing the images, we first identified anomalous (hot) pixels. We started by collecting 100 dark frames of the same exposure time as the data. Setting a threshold equal to 2 standard deviations above the median value of the 100 dark frames, we marked all pixels that are above the threshold in 10\% or more of the frames as hot. These pixels were excluded in further processing. For the data described below, 0.47\% of the pixels were marked as hot. A master dark frame was compiled by taking the mean of the dark frames. The master dark frame was subtracted from each of the {x-ray} frames.

The dark-subtracted {x-ray} frame is then scanned. For each pixel above the event threshold, a surrounding 5$\times$5 grid is defined. The set of pixels above the split pixel threshold (`hit pixels') that are connected to the central pixel by a chain of adjoining hit pixels are included in the sum used to estimate the photon energy and the position of the central pixel is recorded as the {x-ray} position. The event and split pixel thresholds were optimized by examining pulse height spectra for {x-ray}s with energies between 250~eV and 1700~eV (see Sec.~3). The thresholds were adjusted to minimize the width and skew of the pulse height distributions. We found an optimal event threshold of 26~ADU and split pixel threshold of 5~ADU.

\section{Argonne Data}
\label{sect:sections}

\begin{figure}[tb]
\vspace*{5pt}
\centering
\begin{subfigure}{.5\textwidth}
  \centering
  \includegraphics[width=.95\linewidth]{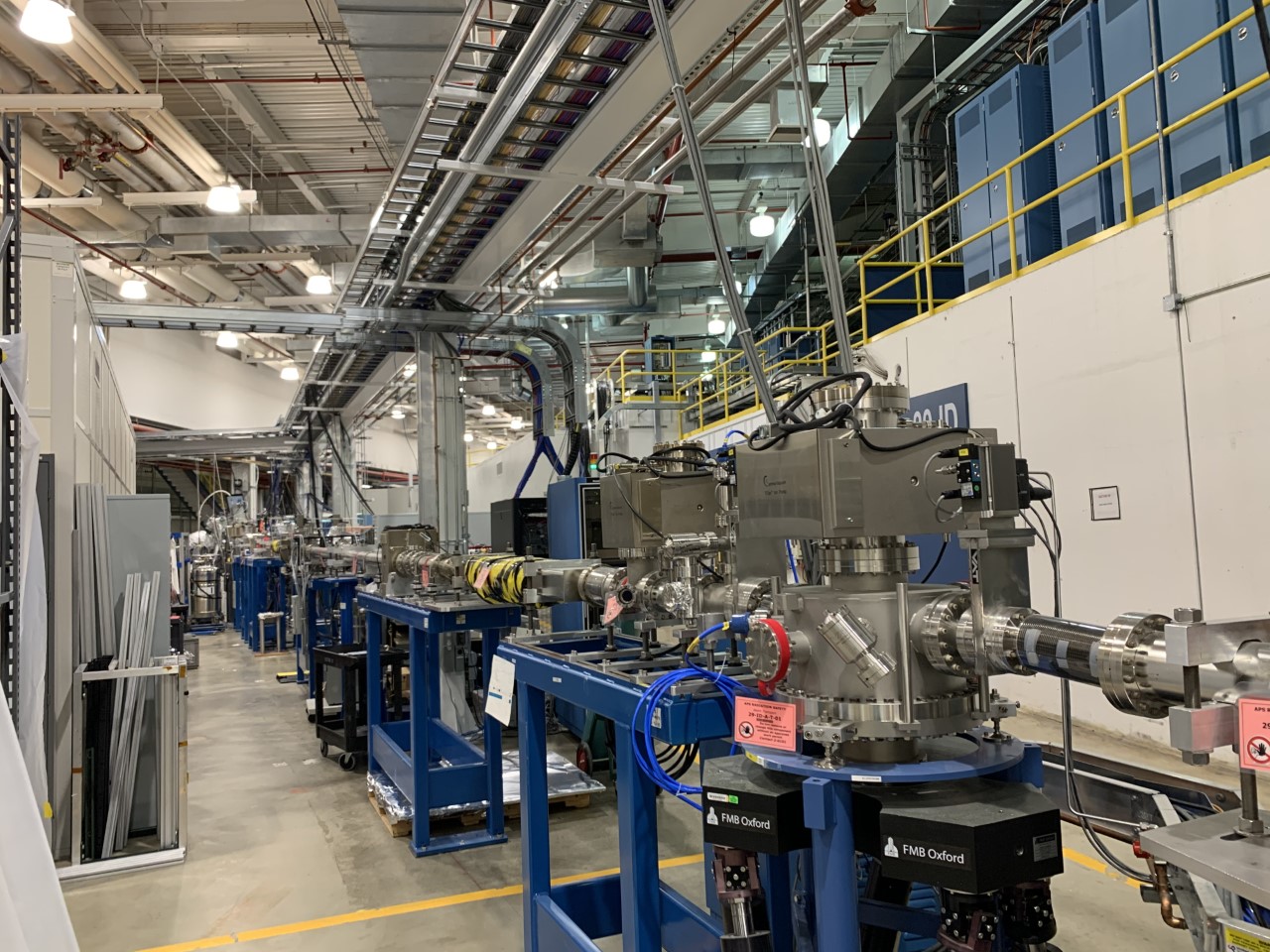}
  \caption{}
  \label{fig:3a}
\end{subfigure}%
\begin{subfigure}{.5\textwidth}
  \centering
  \includegraphics[width=7cm,height=6cm]{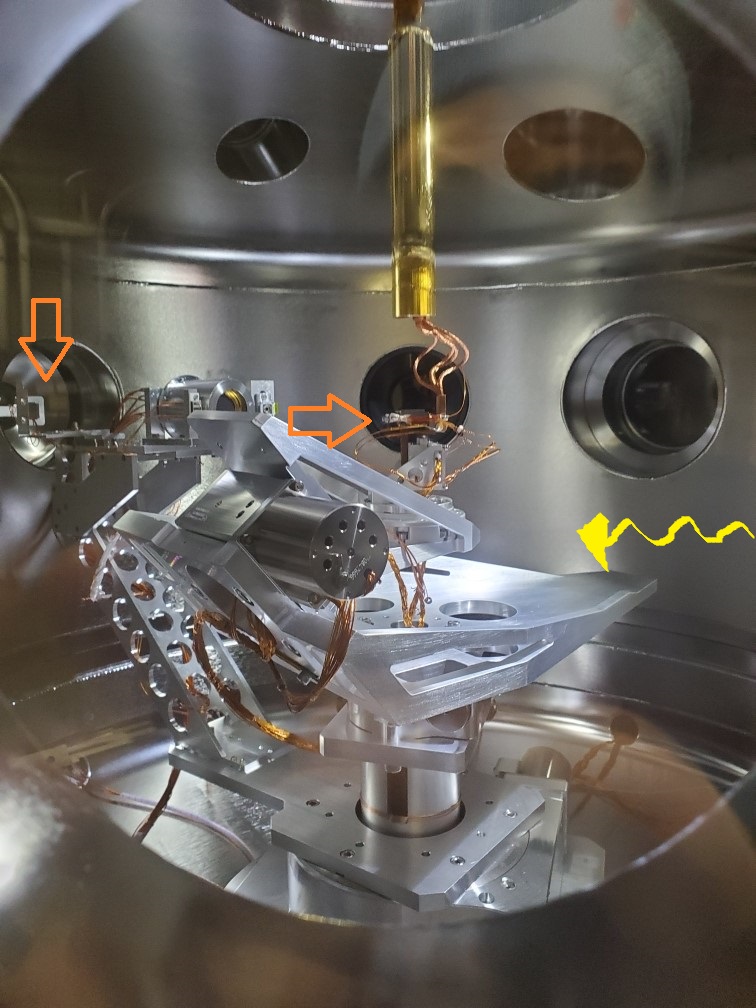}
  \caption{}
  \label{fig:3b}
\end{subfigure}
\caption{(Left) The beamline used at APS for the experiment. (Right) Vacuum chamber where the sensor was held. Arrows from left to right show the sensor port, Si wafer, and beam direction.}
\label{fig:test}
\end{figure}

\subsection{Experimental Setup}

In December 2019, we tested the sensor at the Intermediate Energy {x}-ray (IEX) beamline 29-ID of the Advanced Photon Source (APS) at the Argonne National Laboratory in Illinois. This beamline utilizes an electromagnetic, variably polarizing undulator source followed by a monochromator and is capable of producing monochromatic x-rays between 250-2500 eV\cite{mcchesney2014intermediate}.  Images of the experimental setup are shown in Fig \ref{fig:3a} and Fig \ref{fig:3b}. Our sensor is not capable of handling the intensity of the direct beam in photon counting mode, so it was mounted on a port at an angle 35$^{\circ}$ with respect to the incident beam. A pump near the sensor maintained a high vacuum ($\approx 10^{-9}$ Torr). A Si wafer was used to reflect the incident beam thereby reducing the beam flux ($\sim 10^{12}~\gamma$/s) to a level that could be handled by the camera ($< 100~\gamma$/s). Some experimentation was required to focus the beam in the center of the sensor and the optimal graze angle was 17.5$^{\circ}$.

With the beam centered on the sensor and the vacuum chamber covered to minimize ambient light on the sensor, we accumulated sensor images with an exposure time of 300~ms at energies from 250 eV to 1700 eV (see Table~\ref{tab:energy_scan}). At each energy, we accumulated frames until at least 20,000 photon events were registered. {The sensor maintained an average temperature of  25$^\circ$C during this continuous run.} Both before and after taking beamline data, we turned off the beam and acquired 100 dark frames with 300~ms exposure.

\begin{figure}[tb]
\centering
  \centering
  \includegraphics[width=.85\linewidth]{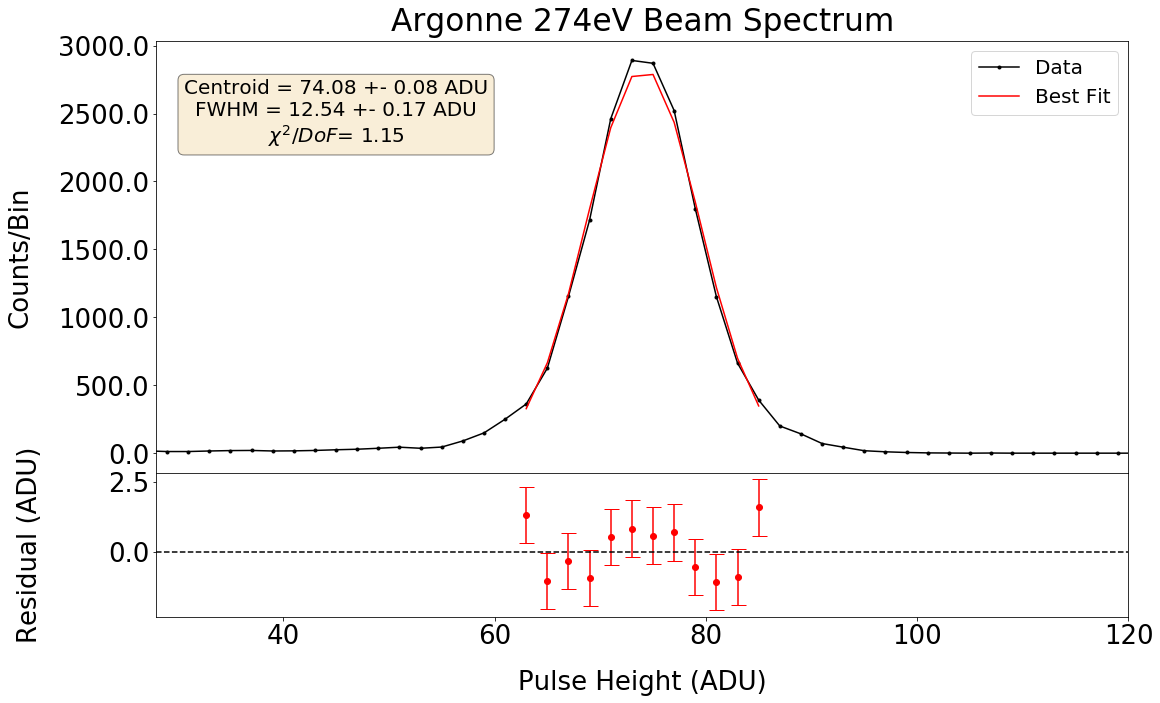}
\caption{{Example of a fitted spectrum from a beamline energy of 274 eV with the best fit Gaussian model.}}
\label{fig:3}
\end{figure}

\subsection{Spectral Fitting}

Using the algorithm described in Sec.~\ref{sec:event_processing}, {x-ray} events were extracted from the sensor images. The list of {x-ray} events was then used to produce an energy spectrum at each beam energy using a bin width of 2 ADU/bin. {We fitted each spectrum with a Gaussian model over a pulse height band covering the peak of the Gaussian, as seen in the solid line in Fig \ref{fig:3}}.

The error on the counts in each pulse height bin was calculated as the quadrature sum of the Gehrels variance function\cite{1986ApJ...303..336G} and a linear term due to differential non-linearity (DNL) proportional to the counts in each bin. The same DNL factor was used for all spectra and was adjusted until the fits produced $\chi^{2}_{\nu} \approx 1$ giving a value of {0.045}. We interpret this as the DNL of the analog-to-digital converters in the IMX290. Table \ref{tab:energy_scan} shows the best fit parameters for each spectrum.

\addtolength{\tabcolsep}{.5em}
\begin{table}[tb]
\vspace*{10pt}
\centering
\scalebox{1.0}{
\begin{tabular}{c c c c c c }
\hline\hline
Energy & Centroid & FWHM &  DoF & $\chi^{2}_{\nu}$ \\ 
 (eV) & (ADU) & (ADU) & &\\
\hline
250.0 & 67.0 $\pm$ 0.1 & 12.4 $\pm$ 0.2 & 8 & 1.7\\
274.0 & 74.1 $\pm$ 0.1 & 12.5 $\pm$ 0.2 & 9 & 1.2\\
280.0 & 74.9 $\pm$ 0.1 & 12.6 $\pm$ 0.2 & 8 & 0.7\\
400.0 & 108.4 $\pm$ 0.1 & 14.7 $\pm$ 0.1 & 10 & 0.5\\
574.0 & 155.0 $\pm$ 0.1 & 15.2 $\pm$ 0.2 & 11 & 1.2\\
950.0 & 257.0 $\pm$ 0.1 & 16.7 $\pm$ 0.1 & 12 & 0.5\\
1200.0 & 325.2 $\pm$ 0.1 & 18.2 $\pm$ 0.1 & 13 & 0.5\\
1450.0 & 392.6 $\pm$ 0.1 & 20.0 $\pm$ 0.2 & 15 & 0.7\\
1700.0 & 459.7 $\pm$ 0.2 & 21.9 $\pm$ 0.4 & 18 & 2.5\\
\hline\hline
\end{tabular}}

\begin{minipage}{1\textwidth}
\vspace*{5pt}
  \caption{{Summary of fit values obtained with a Gaussian plus a constant model.}}
  \label{tab:energy_scan}
\end{minipage}
\end{table}

\subsection{Energy Calibration}

\begin{figure}[tb]
\centering
  \centering
  \includegraphics[width=.85\linewidth]{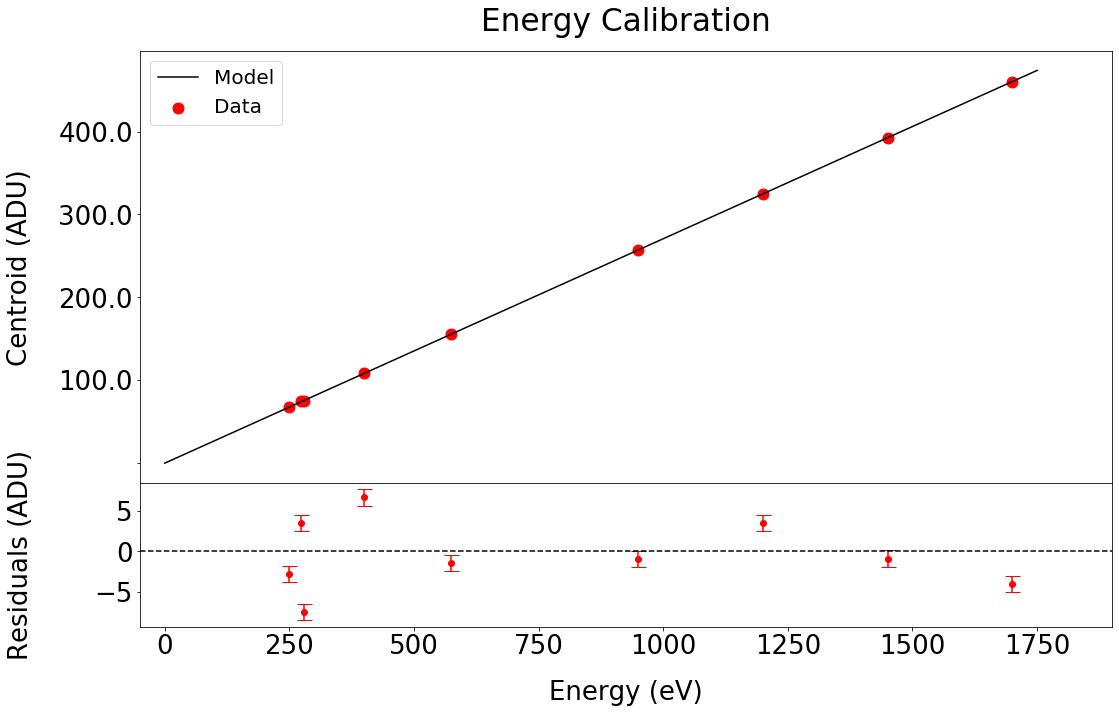}
\caption{{Best fit linear regression to the centroid of the Argonne IEX APS lines with residuals plotted in the lower figure.}}
\label{fig:4}
\end{figure}

We performed a linear regression of the best fitted Gaussian centroids versus beam energy as shown in Fig. \ref{fig:4}. We inverted the relation to find the pulse height to energy calibration. We obtained a best fit slope of {3.6877 $\pm$ 0.0008 eV/ADU} and intercept of {1.840 $\pm$ 0.005 eV} yielding the following conversion function:

\begin{equation}
\label{eq:1}
\text{Energy (eV)} = \frac{{3.69} \text{\, eV}}{\text{ADU}} (\text{ADU value} + {1.8} \ \text{eV}).
  \tag{1}  \
\end{equation}

\noindent The slope is equivalent to a camera gain of {$0.9870 \pm 0.0002 \, {\rm ADU}/e^{-}$}. This is more accurately determined than our initial gain calibration done with the radioactive $^{55}$Fe source and consistent within the uncertainties of that measurement.\\
{With active pixel technology, some variations of gain across the regions of the sensor can be expected. To quantify the variations in gain across the sensor, we divided the sensor area into a 5x5 grid, each grid 387 by 219 pixels. We then calculated the gain within each grid by fitting only the events that occurred within the grid and performing the same linear regression described above. We determined an average gain weighted by the slope error in each grid region, which we find to be 3.686 $\pm$ 0.005 ADU/eV. The gain varies by 0.14\% across the sensor and we choose to use a single gain for the whole sensor.}

\begin{figure}[htb!]
\centering
  \centering
  \includegraphics[width=.85\linewidth]{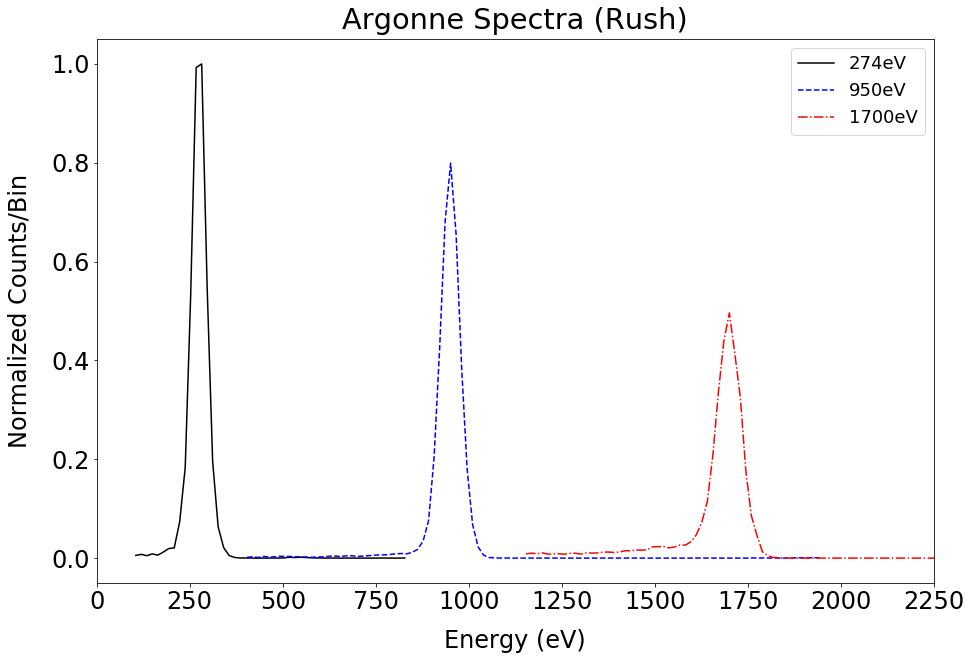}
\caption{{Energy calibrated spectra with a y-axis normalization at beamline energies of 274~eV, 950~eV, and 1700~eV. There are no significant counts below 400~eV and 1150~eV for the 950~eV and 1700~eV lines respectively. We do not plot these spectra below these thresholds for visual clarity.}}
\label{fig:5}
\end{figure}

In Fig. \ref{fig:5} we display calibrated spectra from various beam energies on the same plot. {The Gaussians were normalized to have the same area under their curves.}

\begin{figure}[th]
\centering
  \centering
  \includegraphics[width=.90\linewidth]{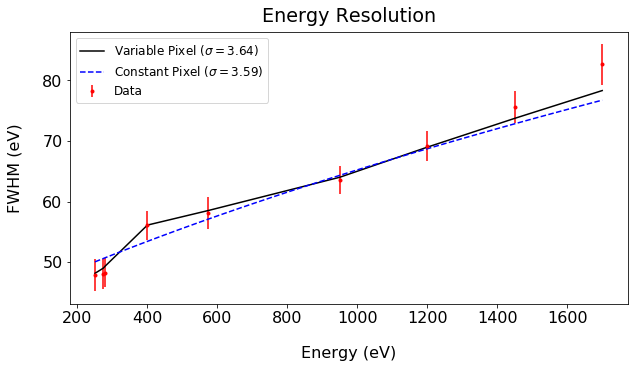}
\caption{Energy resolution (FWHM) modeled by Eq \ref{eq:2}.}
\label{fig:fwhm}
\end{figure}

\subsection{Energy Resolution}

The IEX resolving power is $10^4$; thus, the finite widths of the spectral peaks are due to the sensor. Fig.~\ref{fig:fwhm} shows the energy resolution in terms of the Gaussian full width half max (FWHM) as a function of beam energy. We modeled the energy resolution as the quadrature sum of the Fano limited Poisson fluctuations in the number of photoelectrons produced per photon and the electronic noise per pixel,

\begin{equation}
\label{eq:2}
\text{FWHM} = 2.355 w \; \sqrt{N \sigma^2 + \frac{f E}{w}}  \tag{2}  \
\end{equation}

\noindent where $w = 3.64$ eV is the average ionization energy of silicon, $f
= 0.115$ is the Fano factor for silicon \cite{Owens}, $N$ is the number of pixels used to calculated the pulse height sum for each event, and $\sigma$ is the root-mean-square (RMS) electronic noise per pixel. 

\begin{figure}[th]
\centering
  \centering
  \includegraphics[width=.90\linewidth]{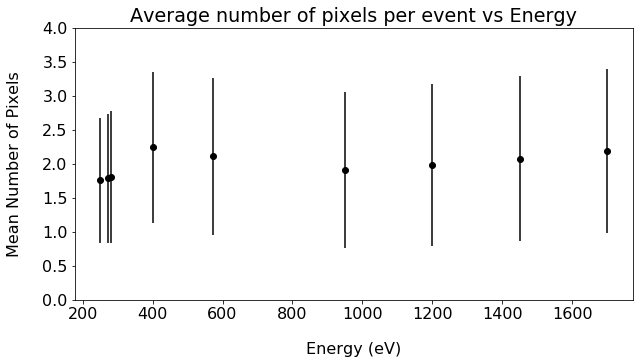}
\caption{Mean number of summed pixels versus energy.}
\label{fig:sumpix}
\end{figure}

Fig.~\ref{fig:sumpix} shows the average number of pixels used in the pulse height summation as a function of energy. The error bars indicate the standard deviation. We attempted two different variations of $N$ for fitting, the first uses a constant median $N$ value of 2 and the second uses the variable mean number of pixels as a function of energy as shown in Fig. \ref{fig:5}. We allowed the electronic noise $\sigma$ parameter to vary and fit the two models to the resolution curve in Fig. \ref{fig:sumpix}. We found best fit values of {$3.64 \pm 0.05 \, e^{-}$} and {$3.59 \pm 0.10 \, e^{-}$} for the variable and constant pixel models, respectively. Our FWHM resolution values of {48.1~eV} at 274~eV and {75.6~eV} at 1450~eV are competitive with the values of $\sim$50~eV and $\sim$70~eV near 0.3~keV and 1.5~keV, respectively, for the back-side illuminated CCDs aboard the \textit{Suzaku} {x-ray} Imaging Spectrometer \cite{bautz2006improved,lamarr2004ground}, and also with the resolution of 96~eV at 1.49~keV reported for the ACIS CCDs \cite{chandra2000chandra}.

{We fitted the Mn~K-$\alpha$ and Mn~K-$\beta$ lines in the spectrum mentioned in Sec \ref{sec:x-ray detection}.  The energy resolution values are larger than those estimated using Eq \ref{eq:2} and the best fit noise for energies below 2 keV. The degradation of the energy resolution above 2 keV may be due to incomplete charge collection. Significant degradation of energy resolution at high energies has been found in a front-illuminated sensor and attributed to a reduction in charge collection efficiency for X-ray that penetrate into the $p^+$ substrate beneath the epitaxial layer \cite{haro2020soft}. This preferentially occurs for higher energy X-rays since they have longer absorption depth. The $p^+$ layer is absent in back-illuminated devices, but incomplete charge collection may occur in the pixel circuitry layer or the carrier wafer.}

\begin{figure}[tb]
\centering
  \centering
  \includegraphics[width=.90\linewidth]{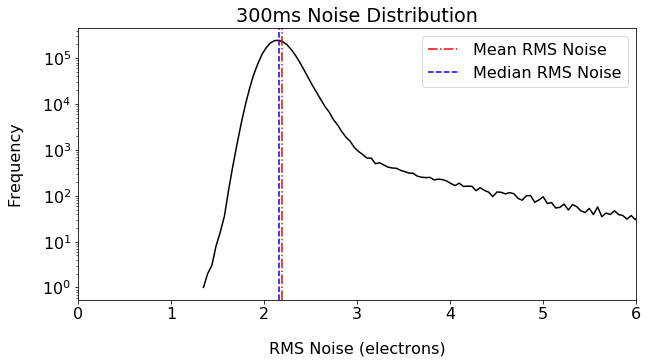}
\caption{RMS noise distribution of the IMX290LLR chip at 300 ms. The vertically and non-vertically dashed lines correspond to the mean and median of the distribution respectively.}
\label{fig:noise}
\end{figure}

\section{Noise Measurements}


The sensor noise can be measured by examining the average variance in each pixel when a sensor is subject to no light at short exposures. Due to the construction of the CMOS pixel, namely that each pixel contains its own circuitry, the RMS noise varies between pixels. This is different than a CCD detector which traditionally has a single amplifier and analog-to-digital converter for all pixels and therefore maintains uniform noise for all pixels.

We evaluated the noise of the IMX290 by taking a series of 100 frames with the sensor in the dark in a vacuum chamber. Each image had an exposure 300 ms and the sensor was held at a constant temperature of 21$^\circ$C. We constructed a master dark frame by taking the mean ADU value in each pixel of the 100 images and then subsequently subtracting the master image from each individual one.  

We constructed a distribution of the per pixel noise by taking the RMS of the 100 master subtracted images (Fig.~\ref{fig:noise}). We find mean and median read noise values of $2.18 \, e^{-}$ and $2.17 \, e^{-}$, respectively. The tail extending to large read noise of the distribution is well known for CMOS sensors and has been characterized as Random Telegraph Signal (RTS) noise. RTS is generated from defects within the silicon that lead to increased electron entrapment and signal variance upon readout.\cite{martin2009rts} Fewer than 0.2\% of the pixels have an RMS noise greater than $4 \, e^{-}$.

\begin{figure}[tb]
\vspace*{5pt}
\centering
\begin{subfigure}{.5\textwidth}
  \centering
  \includegraphics[width=.95\linewidth]{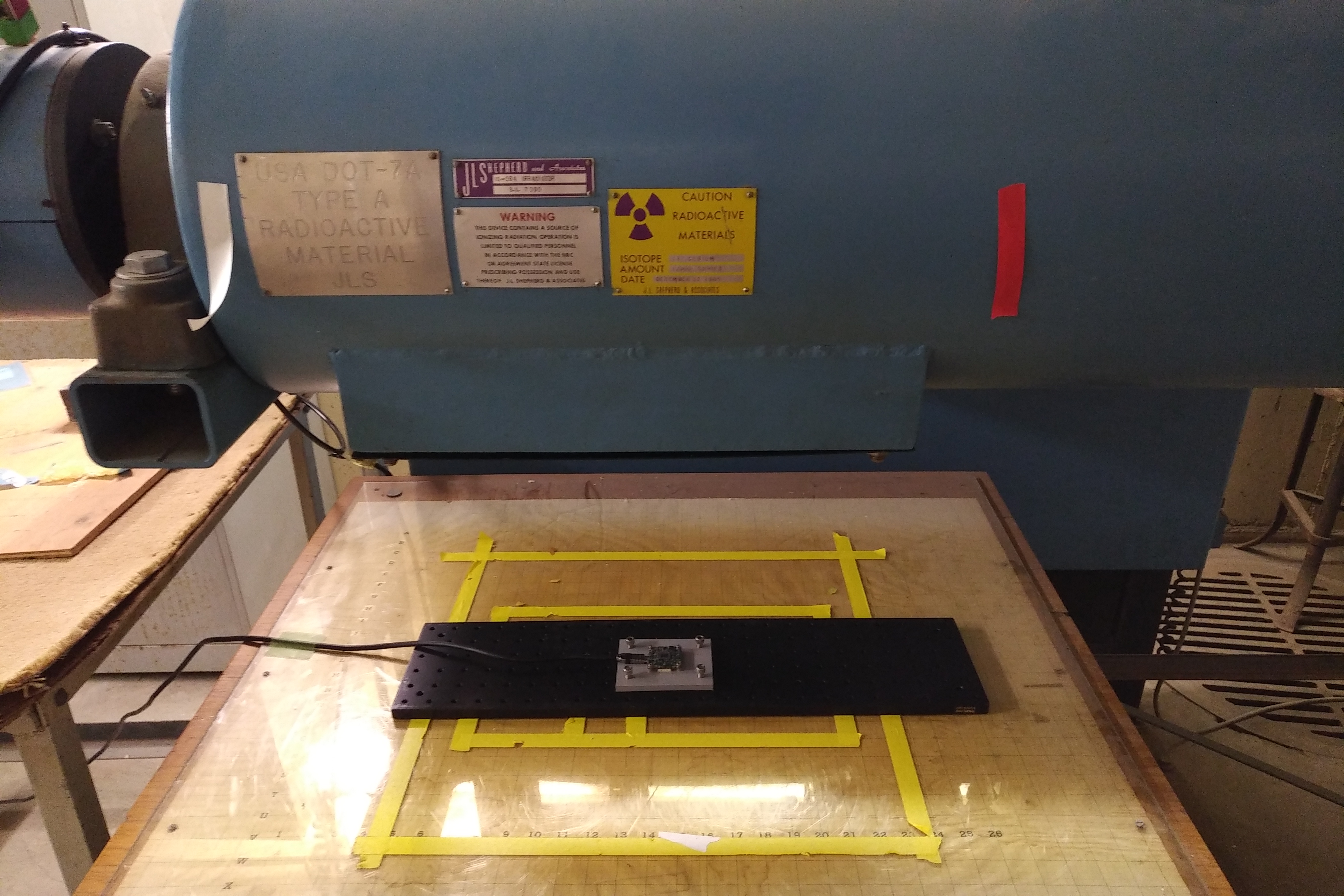}
  \caption{}
  \label{fig:10a}
\end{subfigure}%
\begin{subfigure}{.5\textwidth}
  \centering
  \includegraphics[width=0.94\linewidth]{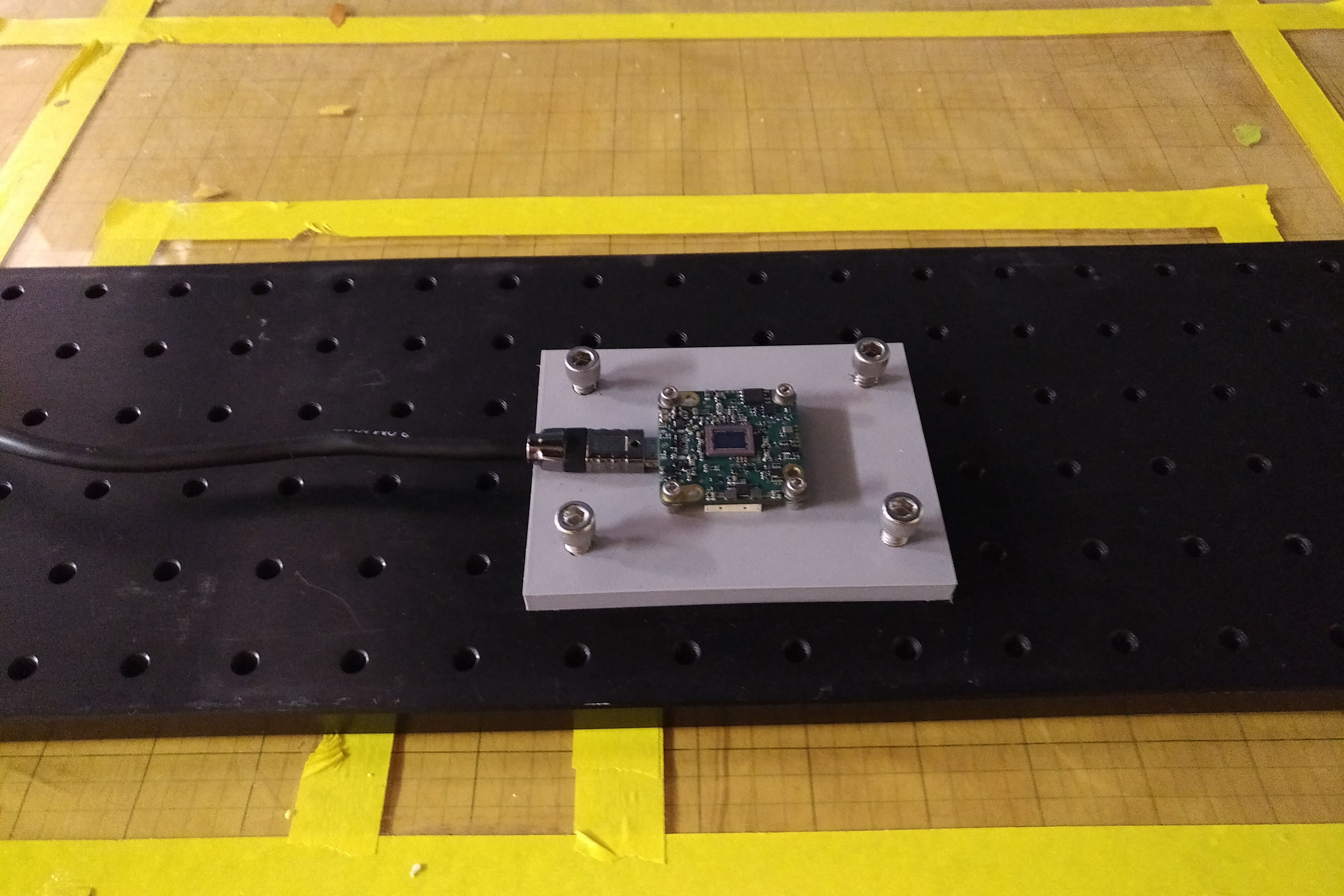}
  \caption{}
  \label{fig:10b}
\end{subfigure}
\caption{The IMX290LLR sensor at RadCORE. (Left) The sensor underneath the Cs-137 sample. (Right) Close up of mount securing sensor}
\label{fig:radcore}
\end{figure}

\section{Radiation Testing}

We irradiated a IMX290LLR sensor with gamma rays to study its performance in a radiation environment. The testing was done at the RadCore facility at the University of Iowa, see Fig.\ref{fig:radcore}. The camera was exposed to gamma-rays from Cs-137 (main peak at 662~keV) at a rate of {0.054} krad(Si)/minute. We operated the camera, recording 10~ms frames at the same gain settings as discussed above, while monitoring the current drawn by the camera via its USB connection from a radiation shielded location. While the camera was operating correctly, the current was 0.16~A. We ended the test after the current dropped to 0.02~A and the camera simultaneously ceased to function. The camera functioned correctly while exposed to radiation for a total of 314.8 minutes, giving a total exposure of 17.1~krad before failure. This is a lower bound on the radiation tolerance of the sensor since some other component on the camera may have failed.

\begin{figure}[tb]
\centering
  \centering
  \includegraphics[width=.90\linewidth]{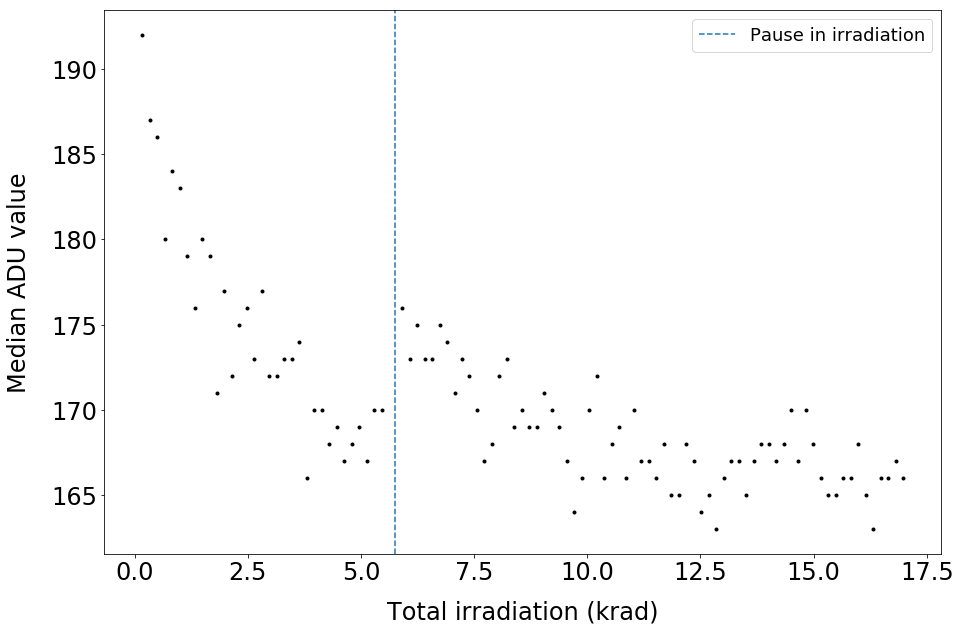}
\caption{Median ADU value versus accumulated dose}
\label{fig:irradiated}
\end{figure}

Fig.~\ref{fig:irradiated} shows the median ADU value versus accumulated dose. The ADU value at the start of the test corresponds to charge deposition of $\sim 20,000 \, e^{-}/{\rm s}$. The median ADU drops by $\approx 15\%$ during accumulation of the first 6~krad, recovers by $\approx 4\%$ during a 13 minute interval during which the irradiation was stopped, and then shows a more gradual decrease during the remaining testing. The variations may be due to ionization damage of the gate oxide in the active pixel circuity.\cite{spieler2008} Such damage anneals at room temperature, consistent with the recovery while irradiation was stopped. Ionization damage depends on rate of irradiation which was much higher during the test than expected in a typical low Earth orbit (LEO), e.g. $\sim$ 5~krad over 2 years. Thus, testing at a low dose rate would be of interest. We note that the sensor continued to operate as the median ADU decreased. Thus, the sensor would likely continue to function, although perhaps with some change in the energy calibration, over a moderate duration mission of several years in LEO.

\section{Conclusion}

We have characterized the performance of a commercial Sony IMX290LLR CMOS sensor as an imaging x-ray spectrometer intended for small satellites. {Radiation testing shows that the sensor and accompanying electronics is operational for well beyond the radiation doses expected for missions of two years duration in low Earth orbit}. Testing at a synchrotron {x-ray} beamline demonstrates the back-illuminated CMOS sensor operated at temperatures near +20$^{\circ}$~C has an energy resolution competitive with the cooled CCD sensors on current x-ray observatories, enabling use of such sensors in the resource-limited environment of small spacecraft.

\subsection* {Acknowledgments}

We thank Amanda Kalen of RadCore for her assistance with the radiation testing. This research was supported in part by the Iowa Space Grant Consortium under NASA Award No. NNX16AL88H. This research used resources of the Advanced Photon Source, a U.S. Department of Energy (DOE) Office of Science User Facility operated for the DOE Office of Science by Argonne National Laboratory under Contract No. DE-AC02-06CH11357; additional support by National Science Foundation under Grant No. DMR-0703406.

\bibliography{report}   
\bibliographystyle{spiejour}   

\end{spacing}
\end{document}